\documentclass{aastex631}

\shorttitle{Hot Subdwarfs identification}
\shortauthors{Tan et al.}

\usepackage{float}
\makeatletter
\let\newfloat\newfloat@ltx
\makeatother
\usepackage{algorithm}
\usepackage{algorithmicx}
\usepackage{algpseudocode}
\usepackage{amsmath}
\usepackage{subfigure}
\usepackage{longtable,booktabs}
\usepackage{tablefootnote}
\usepackage{threeparttable}
\usepackage{appendix}
\usepackage{multirow}

\graphicspath{{./}{figures/}}

\begin{document}

\title{A Robust Hot Subdwarfs Identification Method Based on Deep Learning}

\correspondingauthor{Ying Mei}
\email{meiying@gzhu.edu.cn}

\author[0000-0001-6215-9242]{Lei Tan}
\affil{Center For Astrophysics, Guangzhou University,Guangzhou, Guangdong, China, 510006}
\affil{Great Bay Center, National Astronomical Data Center, Guangzhou, Guangdong, China, 510006}

\author[0000-0002-7960-9251]{Ying Mei}
\affil{Center For Astrophysics, Guangzhou University,Guangzhou, Guangdong, China, 510006}
\affil{Great Bay Center, National Astronomical Data Center, Guangzhou, Guangdong, China, 510006}

\author[0000-0002-1802-6917]{Zhicun Liu}
\affil{CAS Key Laboratory of Optical Astronomy, National Astronomical Observatories, Chinese Academy of Sciences, Beijing, China, 100101}
\affil{University of Chinese Academy of Sciences, Beijing, China, 100049}

\author[0000-0003-3736-6076]{Yangping Luo}
\affiliation{Department of Astronomy, China West Normal University, Nanchong, China, 637002}

\author[0000-0002-8765-3906]{Hui Deng}
\affil{Center For Astrophysics, Guangzhou University,Guangzhou, Guangdong, China, 510006}
\affil{Great Bay Center, National Astronomical Data Center, Guangzhou, Guangdong, China, 510006}

\author[0000-0002-9847-7805]{Feng Wang}
\affil{Center For Astrophysics, Guangzhou University,Guangzhou, Guangdong, China, 510006}
\affil{Great Bay Center, National Astronomical Data Center, Guangzhou, Guangdong, China, 510006}
\affil{University of Chinese Academy of Sciences,Beijing, China, 100049}

\author{Linhua Deng}
\affil{Yunnan Observatories, Chinese Academy of Sciences, Kunming, China, 650011}
\affil{University of Chinese Academy of Sciences, Beijing, China, 100049}

\author[0000-0002-1802-6917]{Chao Liu}
\affil{Key Laboratory of Space Astronomy and Technology, National Astronomical Observatories, Chinese Academy of Sciences, Beijing, China, 100101}
\affil{University of Chinese Academy of Sciences, Beijing, China, 100049}

\begin{abstract}
Hot subdwarf star is a particular type of star that is crucial for studying binary evolution and atmospheric diffusion processes. In recent years, identifying Hot subdwarfs by machine learning methods has become a hot topic, but there are still limitations in automation and accuracy. In this paper, we proposed a robust identification method based on the convolutional neural network (CNN). We first constructed the dataset using the spectral data of LAMOS DR7-V1. We then constructed a hybrid recognition model including an 8-class classification model and a binary classification model. The model achieved an accuracy of 96.17\% on the testing set. To further validate the accuracy of the model, we selected 835 Hot subdwarfs that were not involved in the training process from the identified LAMOST catalog (2428, including repeated observations) as the validation set. An accuracy of 96.05\% was achieved. On this basis, we used the model to filter and classify all 10,640,255 spectra of LAMOST DR7-V1, and obtained a catalog of 2393 Hot subdwarf candidates, of which 2067 have been confirmed. We found 25 new Hot subdwarfs among the remaining candidates by manual validation. The overall accuracy of the model is 87.42\%. Overall, the model presented in this study can effectively identify specific spectra with robust results and high accuracy, and can be further applied to the classification of large-scale spectra and the search of specific targets.

\end{abstract}
\keywords{LAMOST, Hot subdwarf, Spectral Classification, Convolutional Neural Network}

\section{Introduction} \label{sec:intro}
Hot subdwarfs are core helium-burning stars located below the upper main sequence of the Hertzsprung-Russell diagram (HRD) and referred to as extreme horizontal branch (EHB) stars because of their evolution stage. 
In general, Hot subdwarfs are classified into three types by their spectra, i.e.,  subdwarf B (sdB), subdwarf O (sdO), and subdwarf OB (sdOB).

The study of Hot subdwarfs has significant scientific value. \cite{Connell1999} and \cite{2007han} presented that the Hot subdwarfs are the primary sources of the ultraviolet (UV) upturn phenomena found in elliptical galaxies.
They are also critical studies involving close binary interactions because most sdB stars are formed in binary systems. \cite{2007A&AGeier} presented that Hot subdwarf binaries with massive White dwarf companions are believed to be the precursors of type Ia supernova.
The peculiar atmospheres of Hot subdwarfs can be used to study gravitational settling and radiative levitation (\citealt{OToole2006}, \citealt{Geier2013}).
Hot subdwarfs with pulsations can be used for asteroseismic analyses (\citealt{Charpinet2011}).
\cite{2015leii, 2016leii} used Hot subdwarfs to study the extended horizontal branch morphology of globular clusters.
 
Driven by scientific research, searching and identifying Hot subdwarfs and constructing Hot subdwarfs catalogs have become a hot direction in Hot subdwarfs research. The traditional method of searching subdwarfs is mainly based on the basic characteristics of Hot subdwarfs.
For example, \cite{2011ven} identified 48 Hot subdwarfs based on UV photometry of the GALEX survey. \cite{2016luos} identified 166 Hot subdwarfs using photometric information. \cite{2019G} used color, absolute magnitude, and reduced motion cut methods to compile a candidate catalog of 39,800 Hot subdwarfs. However, these identification methods mainly rely on manual processing, which is laborious and difficult to meet the demands of handling large-scale spectral data.

The Large Area Multi-Objective Fiber Optic Spectroscopic Telescope (LAMOST)\citep{cui2012large,zhao2012lamost,deng2012lamost} is a Schmidt telescope with an effective aperture of 3.6$-$4.9m with the field of view of about 5$^{\circ}$. Being a survey telescope, the LAMOST was designed to collect 4000 spectra in a single exposure (spectral resolution R$\sim$1800, limiting magnitude r$\sim$19 mag, wavelength coverage 3700\AA$-$9000\AA).
The LAMOST DR7-V1 catalog contains 10,640,255 calibrated spectra.  The total catalogue includes 9,881,260 STAR spectra, 198,393 GALAXY spectra, 66,406 QSO spectra and 494,196 spectra of Unknown-type. 
  
Identifying Hot subdwarfs from the LAMOST catalog has essential research value because the LAMOST can give the spectral characteristics of Hot subdwarfs that reveal details of the formation and evolution of Hot subdwarfs. However, identifying Hot subdwarfs from LAMOST data is a challenging task. The critical reason is that the LAMOST does not have homogeneous color data (\citealt{2015RAAluo}) and therefore traditional methods such as color-cut (\citealt{2019ApJbu}) cannot be used to search for Hot subdwarfs in the LAMOST catalog. Another reason is the massive amount of catalog data.
Scientists have experimented with other methods to identify HOT subdwarfs from the LAMOST catalog. \cite{2021luo} identified 1587 Hot subdwarfs using the spectra released by the LAMOST and the catalog of Hot subdwarf candidates released by Gaia DR2. 
With the development of machine learning techniques, especially the deep learning techniques, 
\cite{2017ApJSbu} used machine learning methods to search for Hot subdwarf candidates from the LAMOST DR1 and obtained 10,000 candidates, and \cite{2019PASJlei} further identified 56 Hot subdwarfs from these candidates. 
\cite{2019ApJbu} applied Convolutional Neural Network (CNN) to construct a binary classification model for Hot subdwarfs searching in LAMOST DR4 and achieved an F1 value (defined in Section \ref{3.3}) of 76.98$\%$.

Overall, these artificial intelligence-based identification methods have achieved remarkable results. Identifying Hot subdwarf based on deep learning can significantly reduce the difficulty of manual identification and obtain credible results. It can be considered one of the most effective methods to search for a specific target in a large amount of catalog data.

However, there are still some minor deficiencies in current research on deep learning-based recognition, mainly focusing on the imbalance of data samples. 
In the model of \cite{2019ApJbu}, 510 Hot subdwarfs and 5458 other classes (star, galaxy, unknown object, and so on) were used for the binary classification model. From a practical point of view, this ratio is grossly unbalanced. 
Studies have shown that the unbalanced number of datasets for different classes can affect the reliability and accuracy of the classification results. Insufficient sample size can lead to overfitting of the model (\citealt{2018zhan, 2018zhu}). At the same time, the model will be incompleteness due to insufficient sample size (\citealt{2016du}).

In light of the previous literature, a hybrid model with an 8-class classification model and a binary classification model is proposed and applied to search for Hot subdwarfs in the LAMOST catalog.
We introduce the data and preprocessing in Section \ref{2}, including data preparation and data augmentation. In Section \ref{3}, we describe the construction process of our classification model and give a detailed performance evaluation.  In Section \ref{4}, the model is applied to classify the entire LAMOST spectral data to obtain Hot subdwarf candidates. The candidates are identified by manual authentication. The ability of our model is further discussed in Section \ref{5}. In Section \ref{6}, we summarize the work and propose some future work.

\section{Data and Preprocessing}\label{2}
\subsection{Sample set}\label{2.1} 

Based on the LAMOST DR7-V1 catalog,  we created a sample set that extracted the data with all spectra classes (O, Hot subdwarf, White dwarf, B, A, F, G, K, and M) to meet the labeling requirements of the deep learning techniques. 
Referring to the sample selection criteria of \cite{2019ApJbu}, we selected samples with SNR greater than 10 to construct the sample set.
When the SNR was set to be greater than 10, there were sufficient samples for Hot subdwarf, White dwarf, B, A, F, G, K, and M stars, while there were only 72 O-type stars.

\subsection{Data Augmentation}\label{2.2}
To ensure the balance of the sample, we have to augment O-type stars according to the characteristics of the existing 72 samples.
The spectrum of a star is a blackbody radiation spectrum, which is accompanied by emission and absorption lines. That is to say, the generation of a spectrum needs to satisfy characteristics of both two aspects, such as blackbody radiation spectrum and emission/absorption lines.

1. Blackbody radiation spectrum fitting

The blackbody radiation spectrum of a star is defined by the Planck formula (Equation \ref{black_eq}, a relationship between the flux, the wavelength ($\lambda$) and the temperature ($t$) of the spectrum). To further compensate for the effects of relative flux correction and sky background, we modify the formula to Equation \ref{flux}, where $a$ compensates for effects of relative flux correction, and $b$ is compensated for effects of the skylight background.

\begin{equation}
    flux (\lambda, t)=\frac{8\pi hc}{\lambda ^{5}}\cdot \frac{1}{e^{\frac{hc}{\lambda kt}}-1}
	\label{black_eq}
\end{equation}

\begin{equation}
    flux(\lambda ,t,a,b )=a\cdot\frac{8\pi h c}{\lambda ^{5}}\cdot  \frac{1}{e^{\frac{h c}{\lambda k t}}-1} + b
	\label{flux}
\end{equation}

In order to fit the temperature $t$, $a$ and $b$ in Equation \ref{flux}, in practice, we obtained 3700 flux values with wavelengths from $3700\AA$ to $8672\AA$ by truncating each O-type star spectrum. The emission and absorption lines were removed by median filtering, and the corresponding blackbody radiation was obtained.
The fitting process can be realized by the scipy.optimize.curve\_fit\footnote{https://docs.scipy.org/doc/scipy/reference/generated/scipy.optimize.curve\_fit.html} function in the Python Scipy package. 
For each known O-type star spectrum, optimal values were obtained for the parameters ($t$, $a$, $b$) so that the sum of the squared residuals of $flux_{i} - {flux}'_{i}$ was minimized (i.e., Equation \ref{a} was minimized). 
In Equation \ref{a}, $flux$ is the original blackbody radiation spectrum and ${flux}'$ is the fitted flux value obtained by substituting the fitted parameters into Equation \ref{flux}.

\begin{equation}
    \alpha = \sum_{i=1}^{3700} (flux_{i} - {flux}'_{i})^2
    \label{a}
\end{equation}

\begin{figure}[t]
    \centering
    \subfigure[$\alpha$ = 74]{
        \centering
        \includegraphics[width=0.31\textwidth]{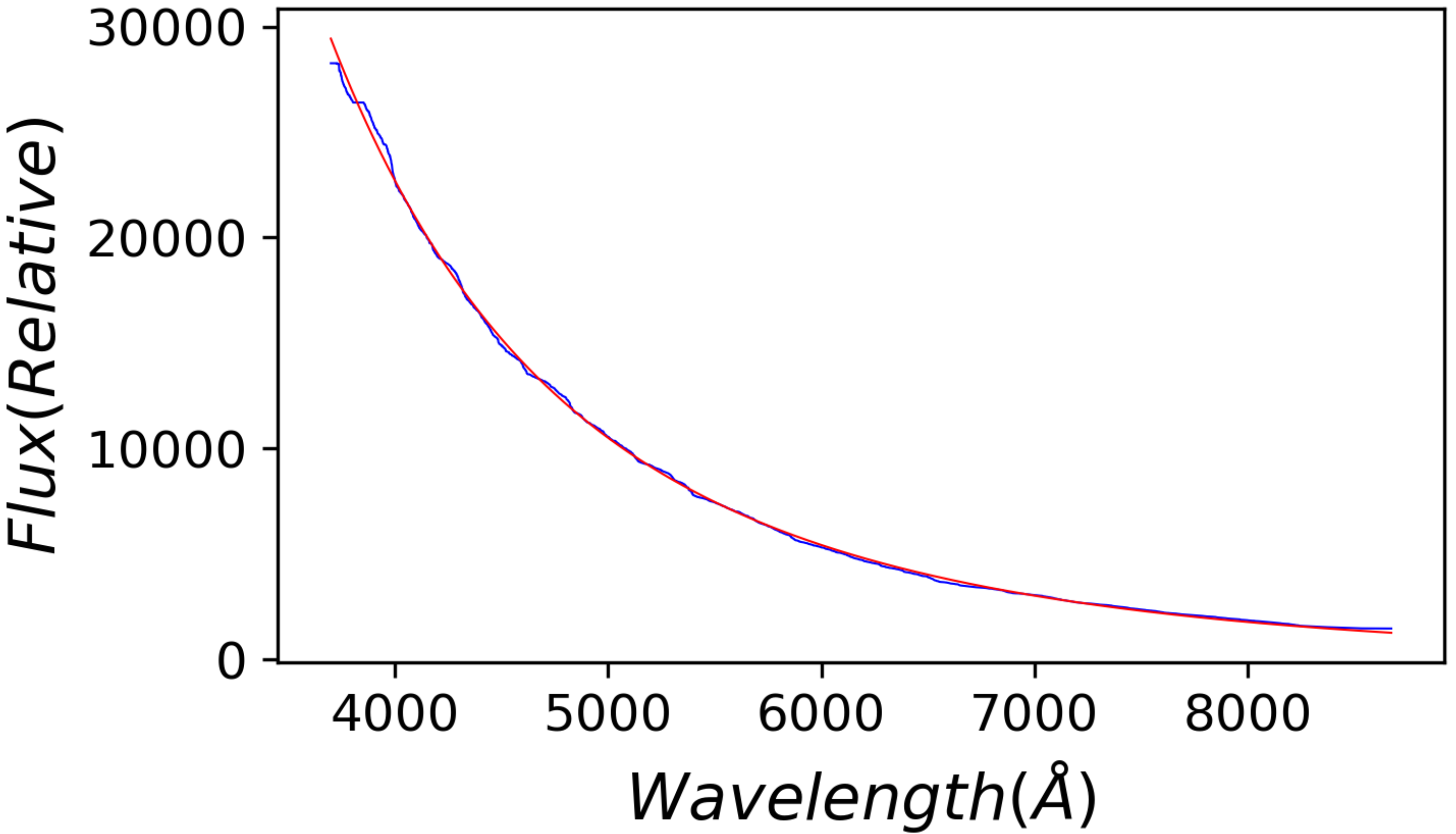}
    }
    \subfigure[$\alpha$ = 146]{
	\includegraphics[width=0.31\textwidth]{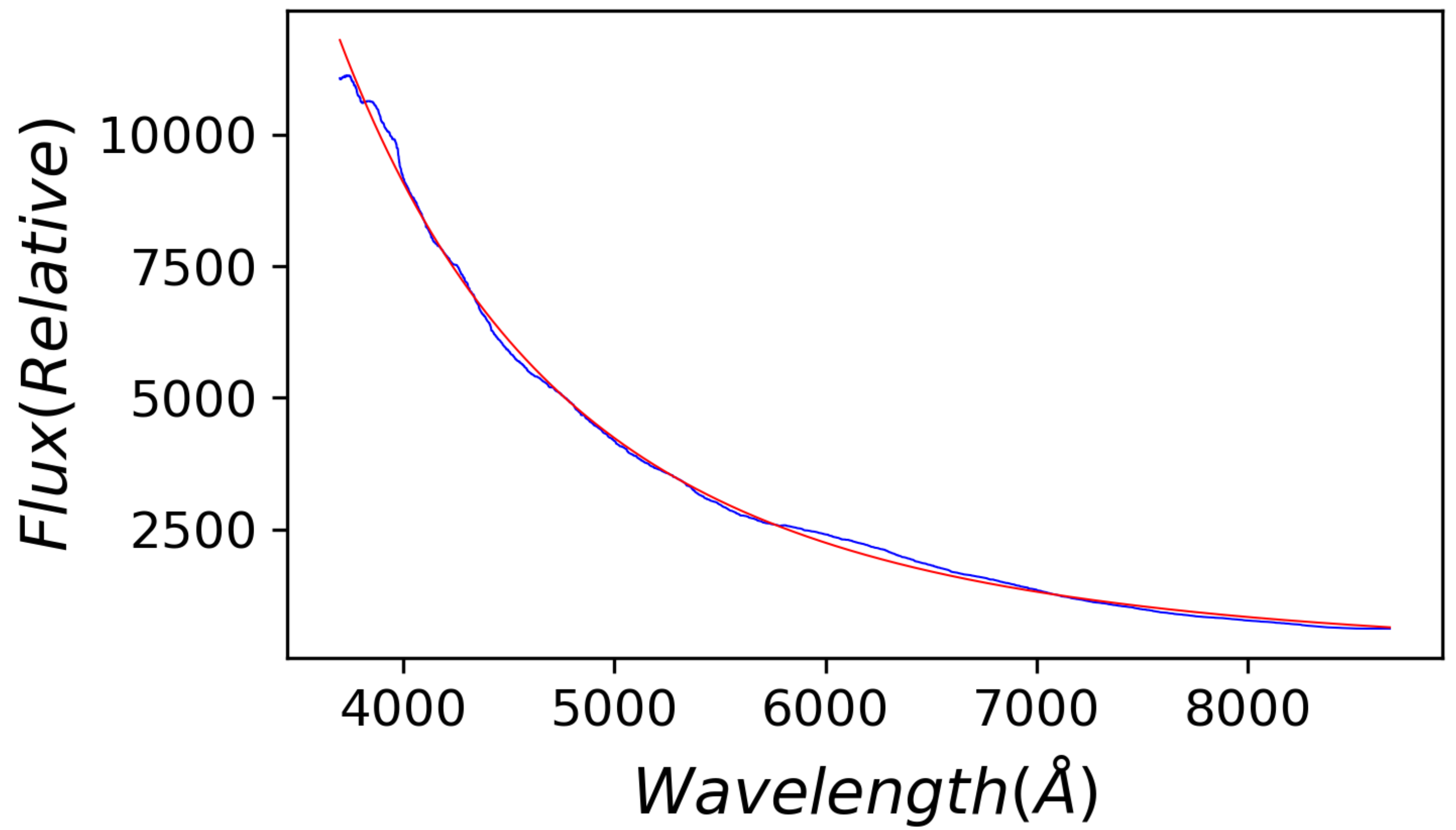}
    }
    \subfigure[$\alpha$ = 502]{
	\includegraphics[width=0.31\textwidth]{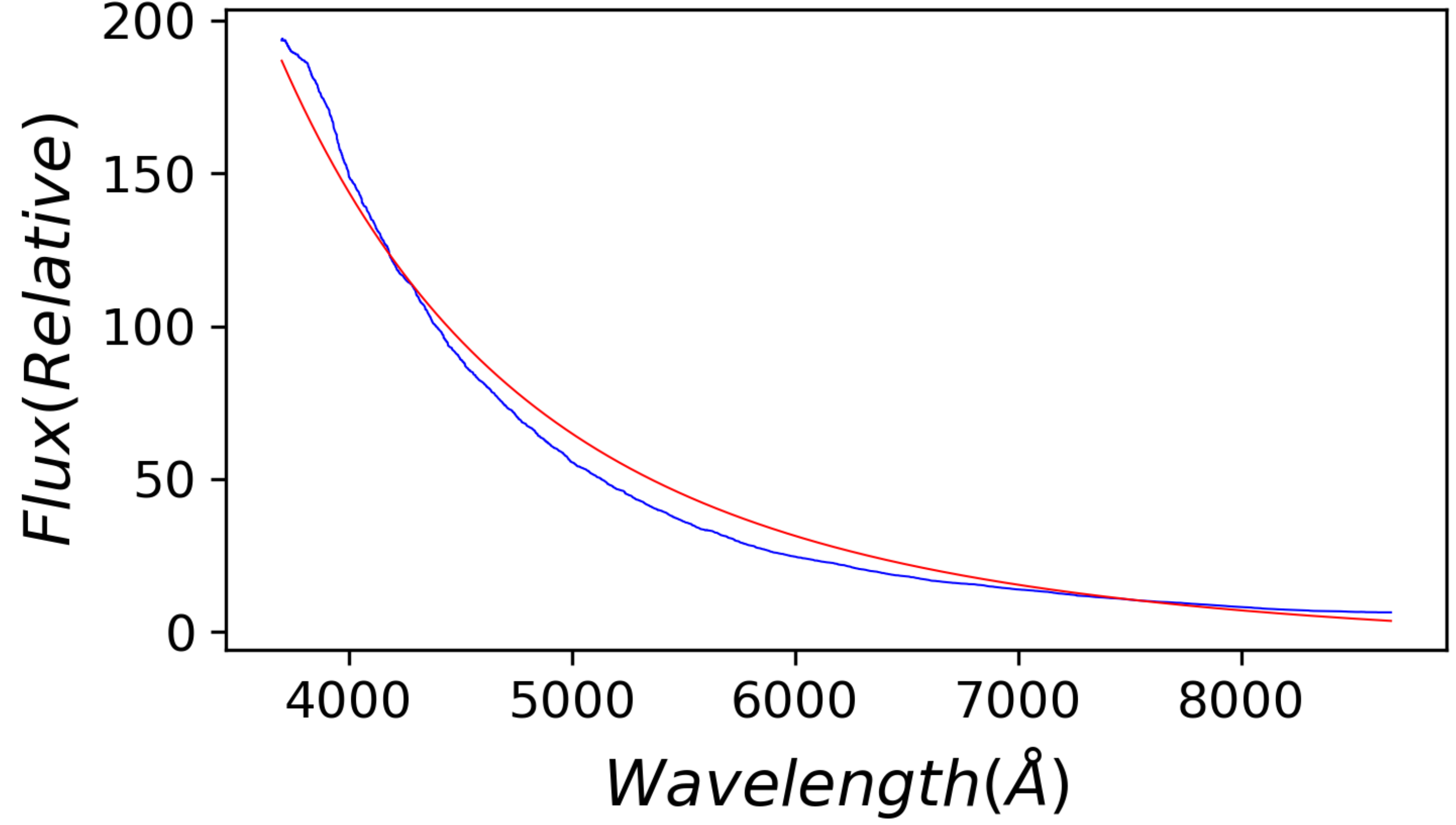}
	}

\caption{The real blackbody radiation spectrum ($flux_{i}$, the blue curve) and the generated blackbody radiation spectrum ($flux'_{i}$, the red curve) under the unified observational conditions (temperature $t$, flux correction $a$ and the sky background $b$). Three samples are given here. The first two are cases where the sum of the squared residuals are relatively small, and sample (c) is a poor fit.}
\label{o_o}
\end{figure}

Consequently, 72 sets of parameters (t, a, b) were obtained by fitting 72 O-type samples.
Figure \ref{o_o} shows the real blackbody radiation spectrum and the generated blackbody radiation spectrum, proving consistent in their physical characteristics.

2. Obtain of emission and absorption lines

Considering that the emission and absorption lines for a specific class are basically the same, the emission and absorption lines of O-type stars can be obtained by subtracting the blackbody radiation spectrum (obtained by median filtering) from the original spectrum. For the 72 known O-type stars, 72 sets of emission and absorption lines can finally be obtained.

To generate a new blackbody radiation spectrum of the O-type star, we set a random temperature (between 28,000K and 100,000K) and pick a random set of parameters $a$, $b$ to obtain the flux calculating by Equation \ref{flux}.
The final generated spectrum was obtained by randomly combining the fitted blackbody radiation spectrum with the emission and absorption lines. 
Algorithm \ref{Sepc_gen} lists the spectrum generation steps in pseudo-code format.From the model training results in section \ref{3.3}, the generated data can solve the overfitting problem caused by the insufficient amount of data well and get a good accuracy rate at the same time, which meets the requirements of deep learning.

\section{Spectral Classification Model Based on CNN}\label{3}
\subsection{Model Design}\label{3.1}
In this section, a CNN-based model for classifying the spectra is proposed. A straightforward idea of model construction is to classify all the spectra into 9 classes and get different types of candidates. However, the preliminary experiments proved that the probability of confusing Hot subdwarfs and White dwarfs in the 9-classification model is high due to their similar blackbody radiation characteristics. 
Therefore, a more refined hybrid model with an 8-class classification model and a binary classification model is constructed. The 8-class classification model disregards the effect of White dwarfs, and the binary classification model further separates Hot subdwarfs from White dwarfs.

The network consists of 6 convolutional layers. A max-pooling layer follows each convolutional layer. After the convolutional and max-pooling layers, two fully connected layers are added. In the convolutional layer, we process the output data by the ReLU function (\citealt{relu}). Cross-entropy function (\citealt{cross}) is used as the loss function, and SGD (\citealt{hardt2016train}) is used as the optimizer. During training, we start with small convolutional kernels and gradually increase the size of the kernels, and the final parameters of each layer are given in Table \ref{modt}. 
The Softmax function (\citealt{ian2016deep}) gives the final classification results, which are probability values of the samples being classified into different classes.

\begin{figure*}[htbp]
\begin{center}
\includegraphics[width=0.8\linewidth]{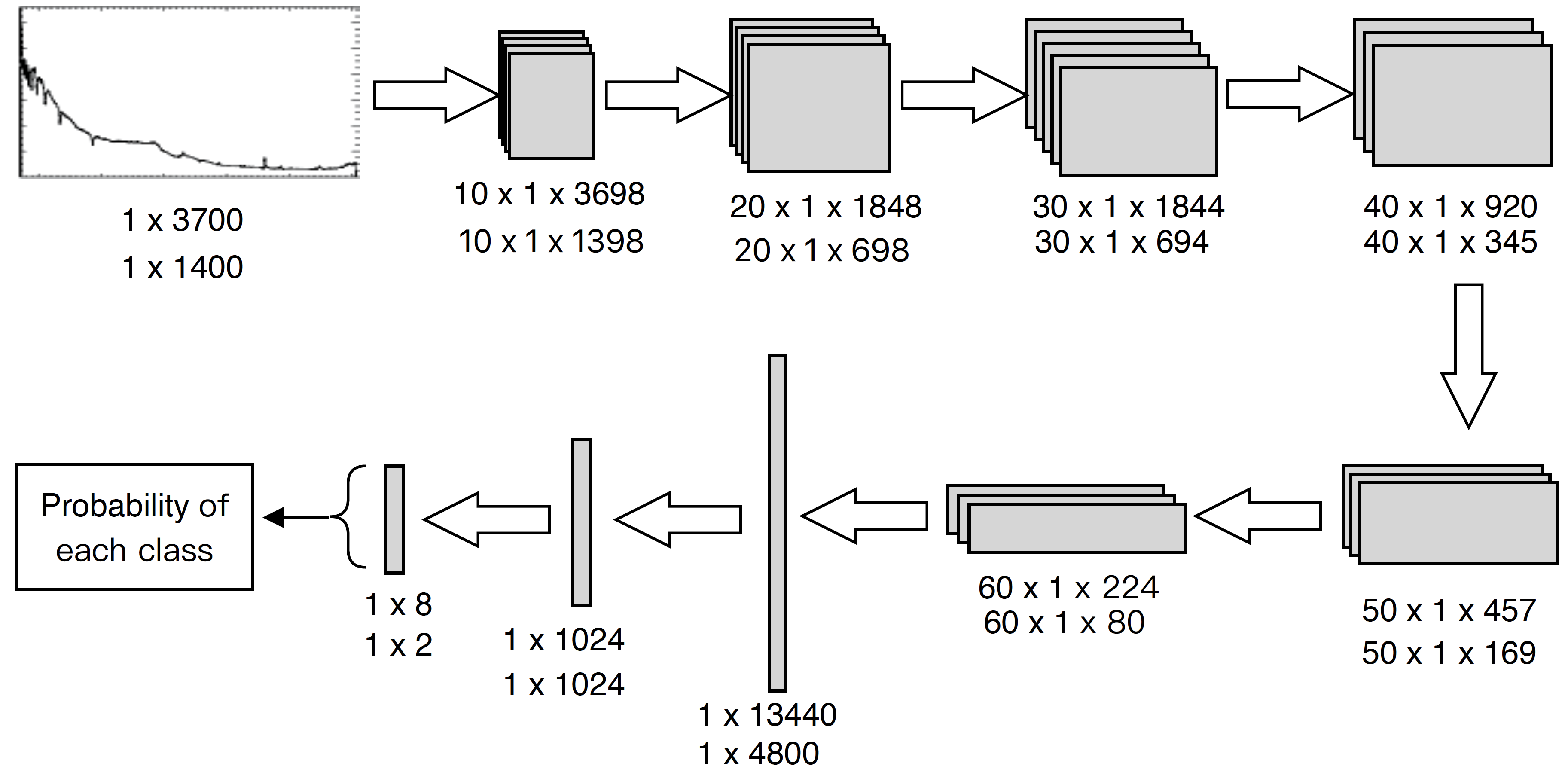}
\end{center}
\caption{The architecture of the classification model. The two rows of numbers under the model represent the feature map size, the top one is for 8-class classification model and the bottom one is for binary classification model.}
\label{mod}
\end{figure*}

\begin{table*}[htbp]
\begin{center}
\caption{Parameters of each layer in the model.}
\centering
\begin{tabular}{cccccc}
\hline
Layer & Type & Maps & Kernel Size & Striding   & Activation\\
\hline
IN & INPUT & 1 & \nodata & \nodata   & \nodata \\
C1 & Convolution & 10 & 1$\times$2 & 1   & ReLU \\
S2 & Max Pooling & 10 & 1$\times$2 & 1    & \nodata \\
C3 & Convolution & 20 & 1$\times$3 & 1    & ReLU \\
S4 & Max Pooling & 20 & 1$\times$2 & 1    & \nodata \\
C5 & Convolution & 30 & 1$\times$4 & 1    & ReLU \\
S6 & Max Pooling & 30 & 1$\times$2 & 2    & \nodata \\
S7 & Convolution & 40 & 1$\times$5 & 1    & ReLU \\
S8 & Max Pooling & 40 & 1$\times$2 & 1    & \nodata \\
C9 & Convolution & 50 & 1$\times$7 & 1    & ReLU \\
S10 & Max Pooling & 50 & 1$\times$2 & 2    & \nodata \\
C11 & Convolution & 60 & 1$\times$9 & 1    & ReLU \\
S12 & Max Pooling & 60 & 1$\times$2 & 2    & \nodata \\
F13 & Fully Connected & \nodata & \nodata & \nodata   & \nodata \\
F14 & Fully Connected & \nodata & \nodata & \nodata   & \nodata \\
OUT & Fully Connected & \nodata & \nodata & \nodata   & Softmax \\
\hline
\label{modt}
\end{tabular}
\end{center}
\end{table*}

\subsection{Data set}\label{3.2}

After sample augmentation of the O-type star spectrum, we randomly selected data and performed manual validation until there were 2000 samples for each class (O, Hot subdwarf, White dwarf, B, A, F, G, K, and M stars, a total of 18,000 samples). 70$\%$ of the samples were randomly selected as the training set, and 30$\%$ were used for testing. 
Considering that our goal was to construct a model for searching Hot subdwarfs, we did not include White dwarfs in the training and testing set of the 8-class classification model. 3700 flux values in wavelength range of 3700\AA \ to 8672\AA \ were obtained by truncating each sample. Specifically, the LAMOST DR7-V1 provides spectral data with a wavelength increment of $\lg (\lambda_{i+1}) - \lg (\lambda_{i}) = 0.0001 $, therefore there are 3700 data points sampled in the wavelength range of 3700\AA \ to 8672\AA \ . We did not resample the original spectral data in our analysis.

The binary classification model was constructed to distinguish Hot subdwarfs from White dwarfs accurately.
Since the blackbody radiation spectrum of Hot subdwarf and White dwarf are very similar, it is difficult for the model to distinguish them when trained with the original spectra. Therefore, for the binary classification model, the emission and absorption lines of the spectra were used for training and testing. 
Experiments showed that the model fit best when trained with the 1400 flux values (the emission and absorption lines) corresponding to wavelengths between 3700\AA \ to 5106\AA \ for each spectrum. 835 Hot subdwarfs with SNR greater than 10 and not involved in the training process were used to further validate the model. The 8-class classification model and binary classification model were trained using the same Hot subdwarf training set to obtain a more robust model.
All of the samples were normalized by Equation \ref{eq:normal} to prevent gradient problems during deep learning model training, where $x_{min}$ was the minimum value of the flux and $x_{max}$ was the maximum value of the flux. 

\begin{equation}
    {x}'=\frac{x - x_{min}}{x_{max} - x_{min}}
	\label{eq:normal}
\end{equation}

\subsection{Modelling and Performance Analysis}\label{3.3}
The performance of a deep learning model is usually evaluated by accuracy, precision, recall, and F1 score (\citealt{forman2003}), which are parameters calculated from the confusion matrix.
A confusion matrix includes four parts, True Positive (TP), False Positive (FP), True Negative (TN), and False Negative (FN).  TP and TN are the observations that are correctly predicted. FP and FN are cases of misclassification. Accordingly, the definition of accuracy, precision, recall and F1 score are in Equation ~\ref{eq:Accuracy}, Equation ~\ref{eq:Precision}, Equation ~\ref{eq:Recall} and Equation ~\ref{eq:f1}.

\begin{figure}[htbp]
    \centering
    \subfigure[8-class classification model]{
        \centering
        \includegraphics[width=0.48\textwidth]{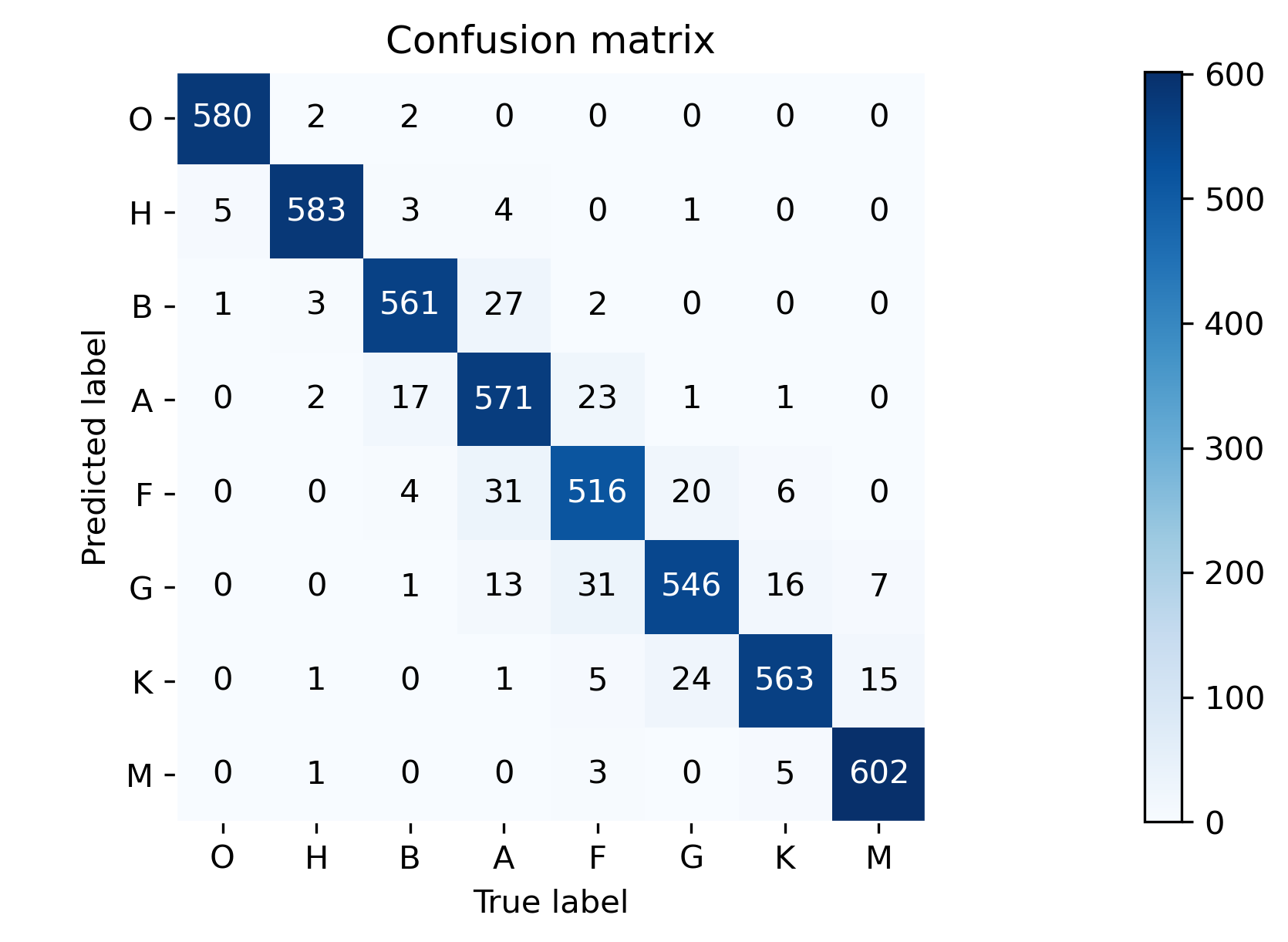}
    }
    \subfigure[Binary classification model]{
	\includegraphics[width=0.48\textwidth]{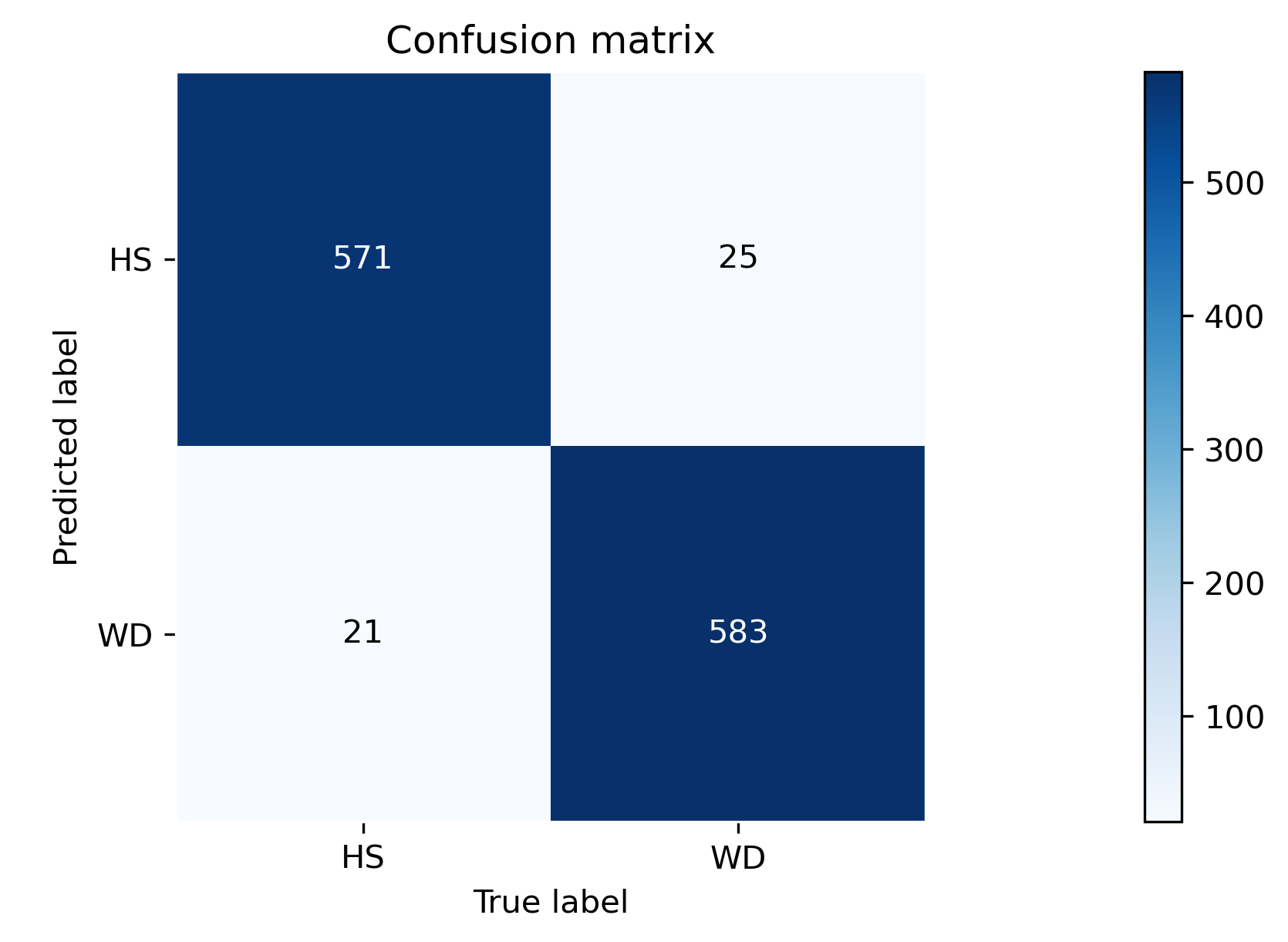}
    }
\caption{The confusion matrix. Subfigure (a) is the confusion matrix of the 8-class classification model and subfigure (b) is the confusion matrix of the binary classification model.}
\label{cm}
\end{figure}

For the 8-class classification model, we set the batch size to 100 and the number of iterations to 2500. After training, we found that the model gradually fitted after about 2200 iterations. For the binary classification model, we set the batch size to 10 and the number of iterations to 2000. The model gradually fitted after about 1750 iterations.
Through repeated training and testing, we finally got an accuracy of 94.21$\%$ for the 8-class classification model and accuracy of 96.17$\%$ for the binary classification model. 
The confusion matrixes are shown in Figure \ref{cm}.
The precision, recall, and F1 score for each class are presented in Table \ref{prf} and Table \ref{brf}. 
We also output the mean $\epsilon$ and variance $\mu$ of the difference between the predicted and true labels of the two models. 
For the 8-class classification model, the mean $\epsilon$ is 0.0166 and the variance $\mu$ is 0.67.
For the binary classification model, the mean $\epsilon$ is 0.023 and the variance $\mu$ is 0.46. The results show that the model performs well, with a tiny difference between the predicted and true labels.

\begin{equation}
    Accuracy = \frac{TP+TN}{TP+FP+FN+TN}
	\label{eq:Accuracy}
\end{equation}

\begin{equation}
    Precision = \frac{TP}{TP+FP}
	\label{eq:Precision}
\end{equation}

\begin{equation}
    Recall = \frac{TP}{TP+FN}
	\label{eq:Recall}
\end{equation}

\begin{equation}
    F1=\frac{2*Precision*Recall}{Precision+Recall}
	\label{eq:f1}
\end{equation}

\begin{table}[htbp]
\centering
\caption{Precision, Recall and F1 score of the 8-class classification model for each class.}
\renewcommand\tabcolsep{22pt}
\begin{tabular}{cccc}
\hline
Class & Precision & Recall & F1 \\
\hline 
O & 98.98$\%$& 99.32 $\%$& 99.15$\%$ \\
H & 98.48$\%$& 97.82 $\%$& 98.15$\%$ \\
B & 95.41$\%$& 94.44 $\%$& 94.92$\%$ \\
A & 88.25$\%$& 92.85 $\%$& 90.49$\%$ \\
F & 88.97$\%$& 89.43 $\%$& 89.20$\%$ \\
G & 92.23$\%$& 88.93 $\%$& 90.55$\%$ \\
K & 95.26$\%$& 92.45 $\%$& 93.83$\%$ \\
M & 96.47$\%$& 98.53 $\%$& 97.49$\%$ \\

\hline
\end{tabular}
\label{prf}
\end{table}

\begin{table}[htbp]
\centering
\caption{Precision, Recall and F1 score of the binary classification model for each class.}
\renewcommand\tabcolsep{20pt}
\begin{tabular}{cccc}
\hline
Class & Precision & Recall & F1 \\
\hline 
Hot subdwarf & 96.45$\%$& 95.81 $\%$& 96.13$\%$ \\
White dwarf & 95.89$\%$& 96.52 $\%$& 96.20$\%$ \\
\hline
\end{tabular}
\label{brf}
\end{table}

To further test the ability of our model for searching Hot subdwarfs, we used the 835 identified Hot subdwarfs which were not involved in the training process for validation. The classification results showed that 802 Hot subdwarfs (96.05$\%$ of the total) were correctly identified. An analysis and discussion of the misclassified samples are given in Section \ref{5}.

\section{Hot subdwarfs Identification and manual Validation}\label{4}
We identified all spectra in the LAMOST DR7 using the hybrid CNN model implemented in the previous section. The Hot subdwarf candidates obtained by the model were verified manually and a final credible catalog was given.

The classification results of the model gave the categories and the corresponding probabilities for all the spectrum of LAMSOT.
By analysing prediction probabilities of the 2235 identified Hot subdwarfs with SNR greater than 10, we found that the model gave a prediction probability of more than 80\% for 2103 of them, wherein the predicted probability of 2043 were more than 99\%. With reference to the work of \cite{Zheng2020}, we considered samples with a prediction probability above 99\% to be more reliable.

In order to obtain reliable results, we did further screening of the results of model predictions from two aspects: (a) For candidates with SNR greater than 10 ($g$-band), we retained the recognized spectra with probability higher than 99\%. (b) For candidates with SNR lower than 10 ($g$-band), we selected the top 200 samples with the highest prediction probabilities. Finally we obtained 2393 Hot subdwarf candidates.

We checked the obtained candidates by cross-validating with the catalog of Hot subdwarfs published by \cite{2021luo} and \cite{2017geier} and manually validating. 
Manual validation results of the 2393 Hot subdwarf candidates are presented in Table \ref{cand}. Along with basic information of the spectra, the last column of Table \ref{cand} describes the manual validation results. The results of manual validation are finally summarized into four categories: 1) Hot subdwarfs that have been identified by previous work, 2) Newly discovered Hot subdwarf stars, 3) Identified as other classes by manual validation, 4) Cannot determine the class due to low SNR.
The results show that among the 2393 candidates, 2092 are Hot subdwarfs (25 are newly discovered), 183 are other types of spectra and misclassified as Hot subdwarfs, and 118 spectra have low SNR and cannot be manually confirmed.
In other words, 87.42$\%$ of the Hot subdwarfs predicted by the model are proved to be Hot subdwarfs in artificial verification.

Following the works of \citet{2016luos} and \citet{2021luo}, atmospheric parameters (Effective temperature T$_{eff}$, surface gravity $log$(g), and $He$ abundances $log(nHe/nH))$ for the 25 newly identified Hot subdwarf stars were obtained by fitting the LAMOST spectra with TLUSTY/SYNSPEC non-LTE synthetic spectra (\citealt{1995hub, 2011hub}). All the parameters of the 25 newly discovered Hot subdwarfs are listed in Table \ref{sdOB}, which includes the LAMOST designation, right ascension, declination, effective temperature, surface gravity, $He$ abundance, the SNR in the $\mu$-band, apparent magnitudes in the $\mu$-band and $g$-band of SDSS DR9, and apparent magnitudes in the $g$-band of Gaia DR2.

\section{Discussions}\label{5}
\subsection{Model performance}\label{5.1}
Compared with the previous binary classification model (Hot subdwarfs and other types), a more accurate hybrid deep learning model is constructed in this paper. It achieved an accuracy rate of 96.17\% on the testing set and 802 of the 835 Hot subdwarfs (96.05\%) in the validation sets are correctly identified.
From Table \ref{brf}, we can see that the F1 value for searching Hot subdwarfs is 96.13$\%$, which is much higher than the 76.98$\%$ in \cite{2019ApJbu}. 

Comparative analysis shows that the overfitting problem caused by insufficient data can be solved by data augmentation. The improvement of training samples also enables the model to learn more complete features, resulting in better results. Overall, the model structure that combined 8-class classification with binary classification model obtains better performance than the binary classification model alone.


\subsection{Misclassification}\label{5.2}
By analyzing the classification results in Section 4, it can be found that for 2428 Hot subdwarfs in the LAMOST catalog, the model eventually found 2067. Therefore, we analyzed the remaining 361 Hot subdwarfs. Through sample analysis, it can be concluded that: (1) 89 Hot subdwarfs were misclassified by our model, (2) 272 Hot subdwarfs were predicted as Hot subdwarfs by our model with a low probability and filtered out during candidate screening (in Section \ref{4}).

\begin{figure}[htbp]
\centering
    \subfigure[spectrum of most Hot subdwarfs]{ 
	\includegraphics[width=0.45\textwidth]{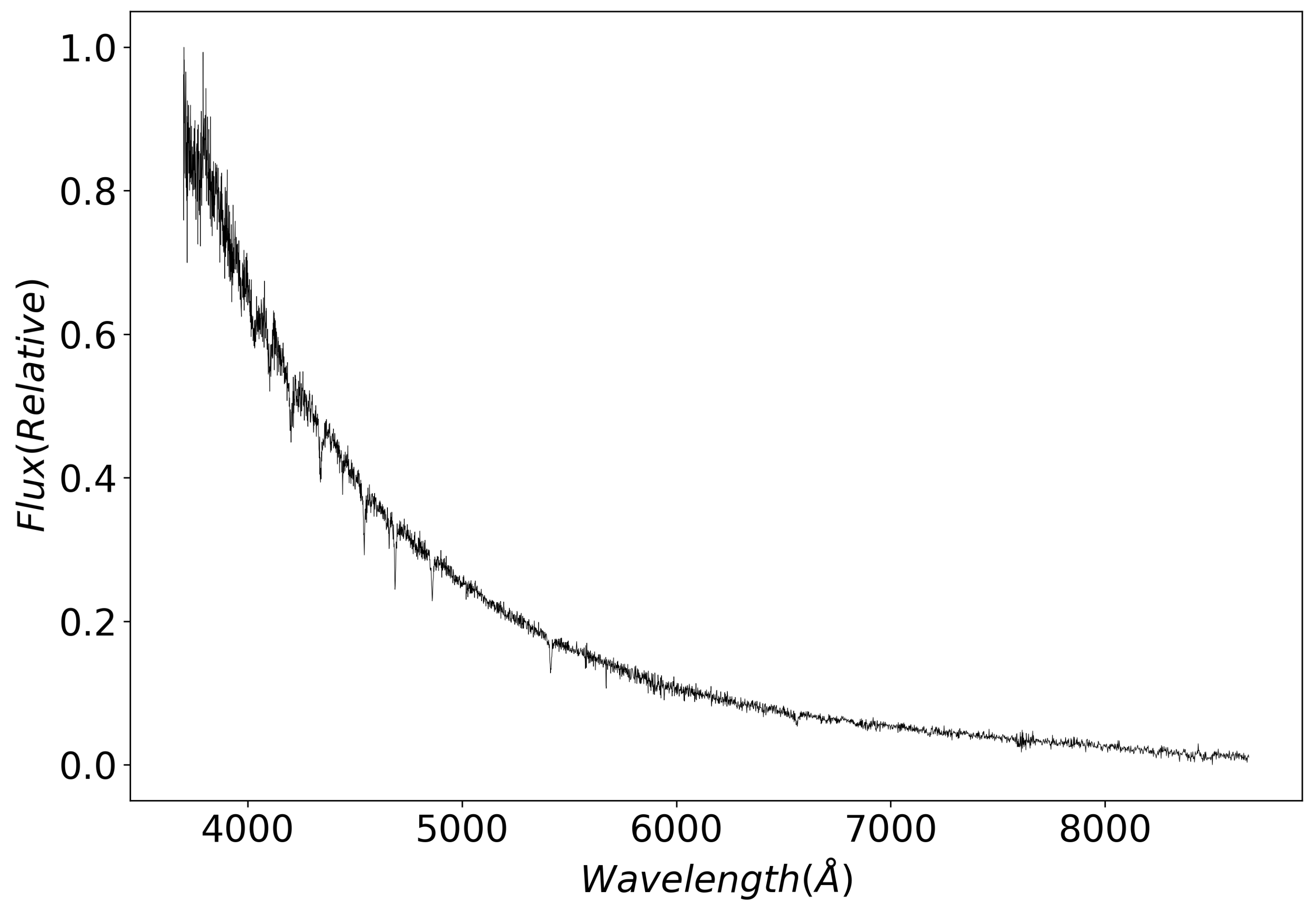}}
    \subfigure[low SNR spectrum]{ 
	\includegraphics[width=0.45\textwidth]{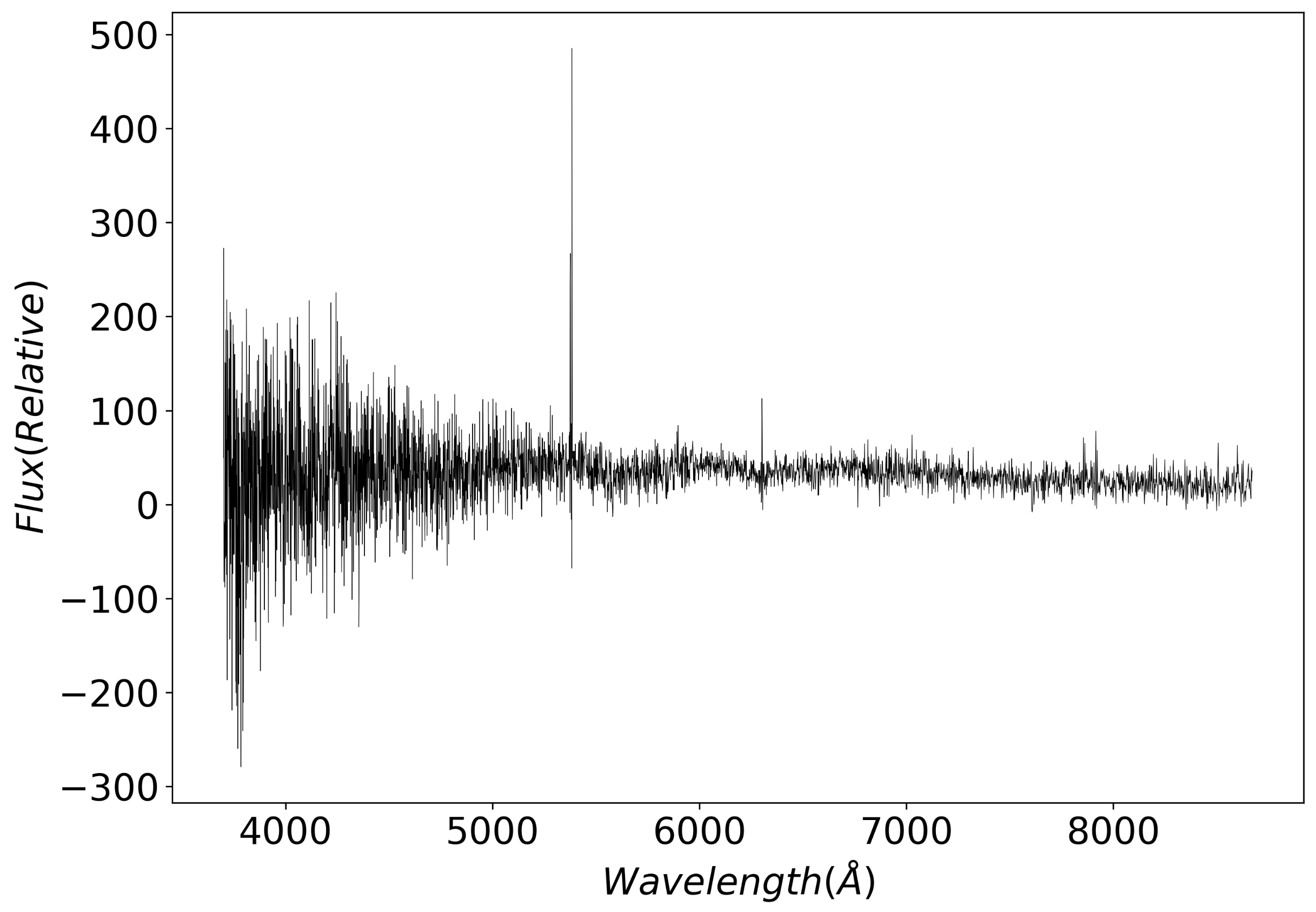}}
    \subfigure[Spectrum with abnormal value]{ 
	\includegraphics[width=0.45\textwidth]{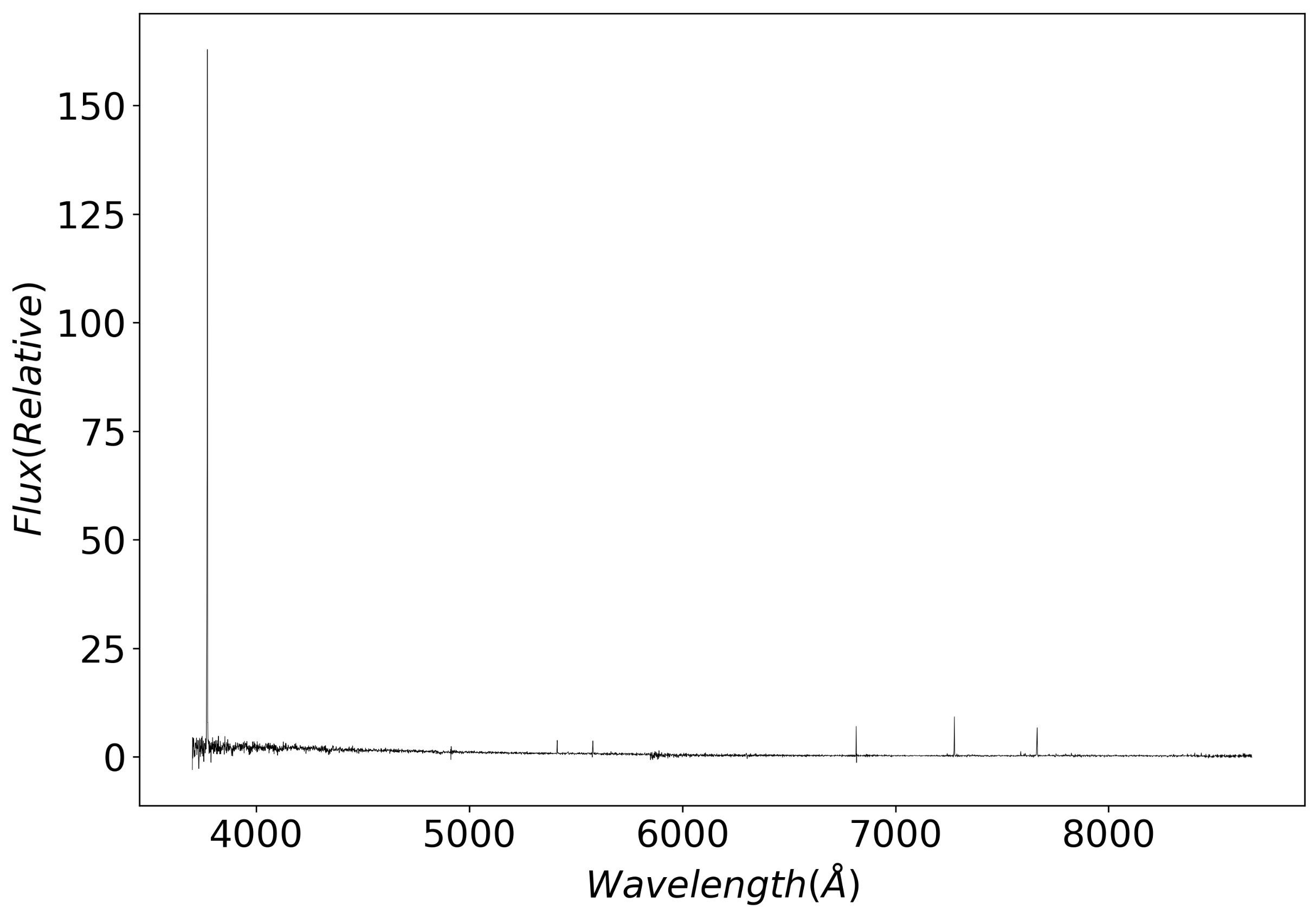}}
    \subfigure[A spectrum exhibits the features of a binary star]{ 
    \includegraphics[width=0.45\textwidth]{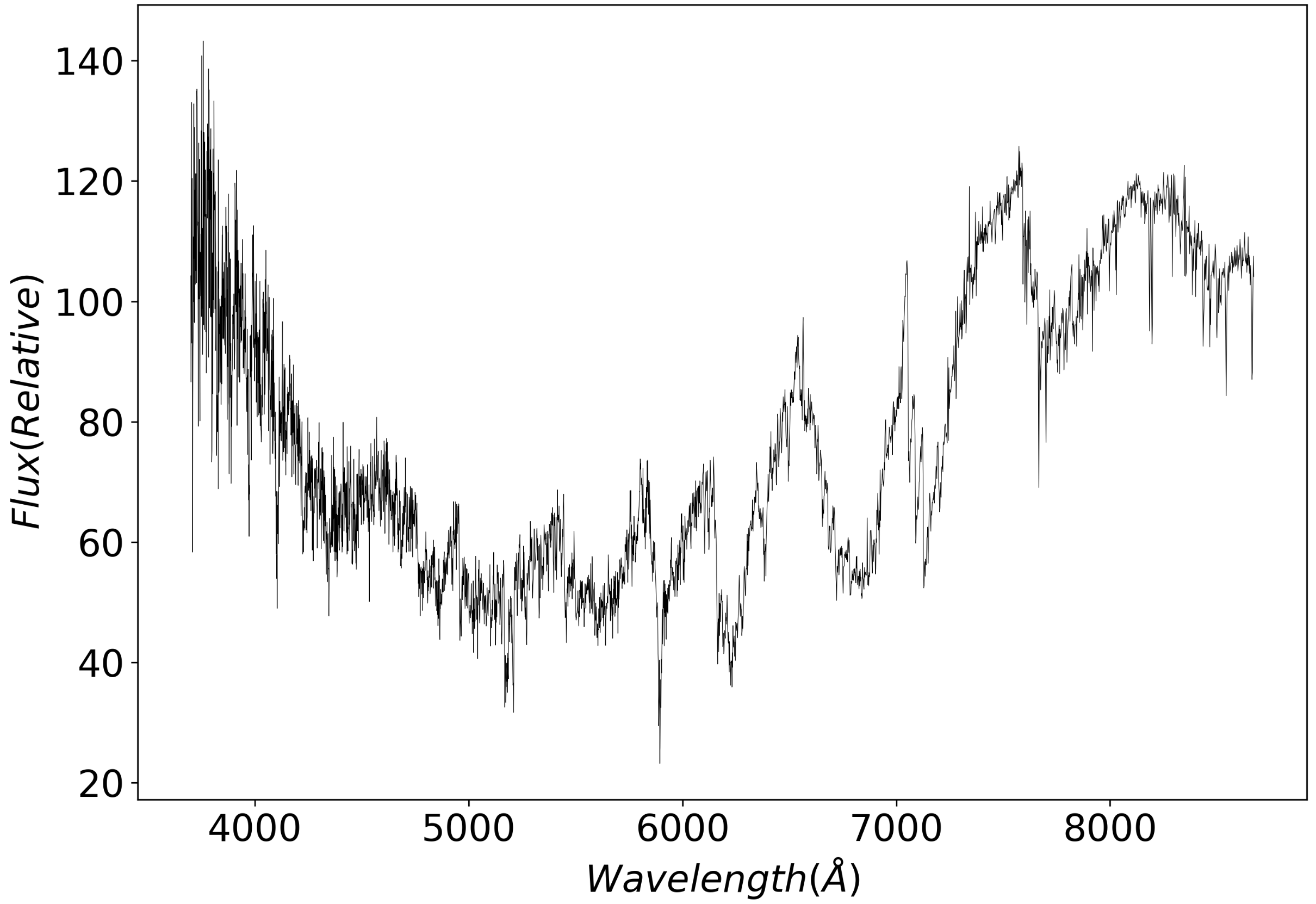}
    }
\caption{Examples of Hot subdwarfs that were misclassified. }
\label{mis}
\end{figure}

The spectral characteristics of most correctly classified Hot subdwarfs are consistent with Figure \ref{mis}a. We performed a detailed analysis of the misclassified spectra and the low probability spectra, and it was found that there are mainly three conditions: (1) Low SNR data (Figure \ref{mis}b). It is difficult to identify the absorption and emission lines. For the effect of noise, our experiments find that the prediction probability gradually decreases as the noise increases. (2) Spectrum with abnormal value, that is, spectrum dominated by sharp spikes either from cosmic rays or instrumental artifacts (Figure \ref{mis}c). We found through experiments that by artificially setting abnormal values for Hot subdwarfs filtered out by our models can also lead to misclassification and low prediction probability. (3) A spectrum exhibits the features of a binary star, as seen in Figure \ref{mis}d, which is probably a binary star (possibly a Hot subdwarf and a M star). To improve the classification accuracy, we need more labeled training data. However, there are very few samples of these types in the LAMOST catalog. These three factors lead to a sample being misclassified or classified with a very low probability as a Hot subdwarf.

For spectrum with low SNR and the category cannot be manually confirmed (Table \ref{cand}, type 4), repeat observations of the LAMOST can be continuously followed at a later stage.

\section{Conclusions}\label{6}

In this study, our proposed approach uses a hybrid CNN-based model to classify the spectra and search for Hot subdwarfs from the LAMOST data. 
For the problem of insufficient data during model training, we propose a spectral generation method based on fitting the Planck formula. 
The model is applied to identify existing Hot subdwarfs in the LAMOST catalog, and most of the Hot Subdwarfs are correctly identified by our model. 
Moreover, our approach offers superior performance to the previous binary classification approach. 
The proposed model is finally used to classify the LAMOST data and search for Hot subdwarfs. The experiments indicate that 87.42\% of the Hot subdwarfs predicted by our model are proved to be Hot subdwarfs. The 25 newly identified Hot subdwarfs would provide a reference for follow-up researches in related fields.

In addition to newly discovered Hot subdwarfs, the current approach is well suited for searching specific targets in vast amounts of spectral data. In particular, spectrum generation provides a solution for samples with a small amount of data and guarantees the reliability of deep learning methods in spectrum classification.
Improving the sample size is one of the following efforts, which is essential for improving the model accuracy. 
We believe that the multi-classification model implemented in this paper is up-and-coming to be successfully applied to search for other specific stars. Capability and performance of the model are worthy of further experiments.
Codes of the spectral generation algorithm, model construction, and model training are available at \url{https://github.com/astronomical-data-processing/cnn_spectra_classification}.

\section{Acknowledgments}\label{7}

This work is supported by the National SKA Program of China (2020SKA0110300), the National Science Foundation for Young Scholars (11903009), the Joint Research Fund in Astronomy (U1831204, U1931141) under cooperative agreement between the National Natural Science Foundation of China (NSFC) and the Chinese Academy of Sciences (CAS), Funds for International Cooperation and Exchange of the National Natural Science Foundation of China (11961141001),
the National Science Foundation of China (12173028), Fundamental and Application Research Project of Guangzhou (202102020677), the Innovation Research for the Postgraduates of Guangzhou University under Grant (2021GDJC-M15).

We thank the anonymous referee for valuable and helpful comments and suggestions. 

\bibliography{bib_cnn_spec}{}
\bibliographystyle{aasjournal}

\renewcommand{\algorithmicrequire}{\textbf{Input:}}  
\renewcommand{\algorithmicensure}{\textbf{Output:}}
\begin{algorithm}[h]
  \caption{: O-type spectrum generation} 
  \label{Sepc_gen}
  \begin{algorithmic}[1]
    \Require 
    \
    \begin{itemize}
      \item $Sepc\_original$: 72 O-type samples, each sepctrum contains 3700 flux values at wavelength from 3700\AA  \ to 8672\AA
      \item $T\_otype$: Temperature of O-type star, $T\_otype\in[28,000K, 100,000K]$
      \item Modified Planck formula (Equation \ref{flux}): $
    flux(\lambda ,t,a,b )=a\cdot\frac{8\pi h c}{\lambda ^{5}}\cdot  \frac{1}{e^{\frac{h c}{\lambda k t}}-1} + b$
	\end{itemize}
	
    \Ensure 
    \
     \begin{itemize}
      \item Generated O-type spectrum, $Sepc\_generation$
     \end{itemize}
    
    \For{each spectrum ($Sepc\_original_{i}$), $i\in [1,72]$}  
      \State /*Step 1: blackbody radiation fitting*/
      \State 1) Perform median filtering on the spectrum and obtain the blackbody radiation spectrum, $flux_{i}$;
      \State 2) According to the modified Planck formula, fit and minimize the value of the objective function $\sum_{i=1}^{3700} (flux_{i} - {flux}'_{i})^2$;
      \State 3) Obtain the corresponding value of the parameters, temperature ($t_{i}$), $a_{i}$ and $b_{i}$.
      \State
      \State /*Step 2: acquisition of absorption and emission lines*/
      \State 4) Obtain the emission and absorption line: $em\_ab\_line_{i} = sepc\_original_{i} - flux_{i}$.
    \EndFor
    \State
    
    \For{$i\in [1, N(number\_of\_spectrum\_to\_be\_generated)]$} 
      \State /*Step 3: O-type spectrum generation*/
      \State 1) Randomly choose a temperature ($T\_otype_{k}$) between 28,000K and 100,000K;
      \State 2) Randomly choose a set of a and b obtained in Step 1, and feed a, b and $T\_otype_{k}$ to Equation \ref{flux} (where $t$ = $T\_otype_{k}$) to generate the blackbody radiation spectrum, $spec\_blackbody\_gen_{k}$;
      \State 3) Randomly choose a set of emission and absorption line obtained ($em\_ab\_line_{i}$) in Step 2;
      \State 4) Generate O-type spectrum: $Sepc\_generation$ = $spec\_blackbody\_gen_{k} + em\_ab\_line_{i}$
    \EndFor
  \end{algorithmic}
\end{algorithm}

\begin{deluxetable}{cccccccccc}
\tablecaption{Manual verification results of the 2393 Hot subdwarf candidates.\label{cand}}
\tablehead{
\colhead{Designation} &  
\colhead{OBSID} &
\colhead{RA$^{\circ\ddagger}$} &
\colhead{Dec$^{\circ}$} &
\colhead{SNR} &  
\colhead{SNR} &
\colhead{SNR} &
\colhead{SNR} &
\colhead{SNR} &
\colhead{ }  \\ 
\colhead{LAMOST} &  
\colhead{LAMOST} &
\colhead{LAMOST} &
\colhead{LAMOST} &
\colhead{$g$-band} &  
\colhead{$\mu$-band} &
\colhead{$z$-band} &
\colhead{$r$-band} &
\colhead{$i$-band} &
\colhead{Type}   
}
\startdata
J023200.24+333436.1 & 632206097 & 38.001041 & 33.576702 & 60.56 & 21.51 & 25.19 & 53.68 & 47.63 & 1  \\
J211531.47+123957.5 & 592502156 & 318.881150 & 12.665982 & 52.55 & 35.69 & 17.17 & 39.60 & 35.41 & 1  \\
J155144.87+002948.8 & 133709100 & 237.936982 & 0.496901 & 18.40 & 7.45 & 6.73 & 18.98 & 15.05 & 1  \\
J071856.25+102638.3 & 446310014 & 109.734390 & 10.443990 & 64.56 & 35.33 & 24.47 & 53.17 & 49.80 & 1  \\
J212356.68+153323.5 & 592415028 & 320.986190 & 15.556550 & 36.19 & 18.30 & 12.39 & 31.63 & 28.95 & 1  \\
J210132.70+135622.5 & 371602229 & 315.386270 & 13.939585 & 37.99 & 29.44 & 10.61 & 27.60 & 22.89 & 2  \\
J034322.46+463222.2 & 399504032 & 55.843622 & 46.539507 & 18.85 & 8.98 & 9.62 & 18.38 & 18.24 & 2  \\
J045031.77+230712.9 & 197203189 & 72.632382 & 23.120270 & 41.74 & 26.74 & 23.16 & 33.79 & 40.42 & 2  \\
J060619.60+200141.7 & 504103064 & 91.581701 & 20.028262 & 43.07 & 16.75 & 36.92 & 40.71 & 50.95 & 2  \\
J182942.24+110429.1 & 746715115 & 277.426030 & 11.074758 & 38.31 & 16.52 & 17.89 & 29.15 & 31.51 & 2  \\
J072222.78+445904.4 & 504804107 & 110.594950 & 44.984569 & 26.37 & 13.70 & 7.60 & 21.90 & 16.67 & 3  \\
J122420.49+264738.8 & 714311122 & 186.085402 & 26.794137 & 17.85 & 8.02 & 7.97 & 13.39 & 13.35 & 3  \\
J113143.40+370128.0 & 450104014 & 172.930840 & 37.024465 & 37.58 & 17.88 & 17.81 & 37.05 & 33.04 & 3  \\
J101134.32+342150.7 & 732201084 & 152.893033 & 34.364110 & 2.92 & 0.85 & 3.17 & 4.18 & 4.56 & 3  \\
J104052.58+284856.7 & 630714012 & 160.219100 & 28.815752 & 21.98 & 7.24 & 7.29 & 14.95 & 13.58 & 3  \\
J222527.75+012522.5 & 484002197 & 336.365666 & 1.422931 & 5.74 & 3.83 & 2.23 & 4.17 & 4.64 & 4  \\
J004511.44+321708.6 & 195305132 & 11.297683 & 32.285725 & 9.69 & 2.65 & 5.83 & 8.55 & 10.90 & 4  \\
J004938.10+460953.1 & 498808130 & 12.408772 & 46.164764 & 5.17 & 1.94 & 6.25 & 6.29 & 9.29 & 4  \\
J095728.68+275506.2 & 50401068 & 149.369516 & 27.918390 & 9.54 & 14.27 & 1.76 & 1.56 & 2.63 & 4  \\
J105916.42+512443.1 & 148611045 & 164.818450 & 51.411984 & 7.49 & 3.89 & 1.52 & 6.28 & 4.22 & 4  \\
...& ...& ...& ...& ...& ...& ... & ...& ...& ... \\
\enddata
\tablecomments{Column 1-9: Basic information in the LAMOST catalog. Column 10: Type description. 1: Hot subdwarfs that have been identified by previous work. 2: Newly discovered Hot subdwarfs. 3: Identified as other classes by manual validation.  4: Cannot determine the class due to low SNR.}

\end{deluxetable}

\begin{table}[htbp]
\centering
\caption{The atmospheric parameters of 25 newly identified Hot subdwarfs in this work.\label{performance}}
\renewcommand\tabcolsep{3.3pt}
\begin{tabular}{cccccccccc}
\hline
Designation† & RA$^{\circ\ddagger}$ & Dec$^{\circ}$ & $T_{eff}$ & $log$(g) & $log$(nHe$/$nH)$^{\S}$ & SNR & uSDSS & gSDSS & G GaiaDR2 \\
LAMOST & LAMOST & LAMOST & (K) & (cm$/$s$^{-2}$) &  & $\mu$-band & (mag) & (mag) & (mag) \\
\hline
J011346.77+002828.6 &  18.444903 &   0.474638 & 66154$\pm$377  & 6.5$\pm$0.03 &  2.38$\pm$0.45 & 63.05 & 14.51 & 14.96 & 15.23 \\
J021334.02+435651.7 &  33.391767 &  43.947714 & 50287$\pm$1961 & 5.86$\pm$0.31 &  -0.07$\pm$0.19 & 2.0 & \nodata & \nodata & 16.89 \\
J023007.26+292256.2 &  37.530274 &  29.382299 & 74994$\pm$1650 & 6.50$\pm$0.08 &  1.89$\pm$2.49 & 21.37 & 16.99 & 17.39 & 17.61 \\
J032425.51+430302.7 &  51.106302 &  43.050760 & 49080$\pm$511 & 6.19$\pm$0.08 &  0.76$\pm$0.13 & 6.28 & 17.52 & 17.58 & 17.67 \\
J034322.46+463222.2 &  55.843622 &  46.539507 & 48674$\pm$346 & 5.81$\pm$0.07 &  1.00$\pm$0.12 & 8.98 & 16.80 & 16.73 & 16.70 \\
J042351.08+322201.0 &  65.962855 &  32.366951 & 50248$\pm$977 & 5.73$\pm$0.13 &  0.11$\pm$0.13 & 3.61 & 17.38 & 17.39 & 17.45 \\
J045031.77+230712.9 &  72.632382 &  23.120270 & 41776$\pm$75  & 5.79$\pm$0.02 &  0.84$\pm$0.02 & 26.74 & 17.73 & 17.56 & 17.47 \\
J051135.26+391706.5 &  77.896933 &  39.285150 & 42434$\pm$273 & 6.31$\pm$0.09 &  1.08$\pm$0.20 & 4.47 & \nodata & \nodata & 17.13 \\
J051304.75+114717.1 &  78.269827 &  11.788084 & 49634$\pm$257 & 6.50$\pm$0.01 &  2.07$\pm$0.11 & 73.99 & \nodata & \nodata & 14.38 \\
J054559.57+222738.0 &  86.498229 &  22.460566 & 46032$\pm$797 & 5.52$\pm$0.11 &  1.03$\pm$0.24 & 8.75 & \nodata & \nodata & 17.63 \\
J060619.60+200141.7 &  91.581701 &  20.028262 & 47105$\pm$321 & 5.87$\pm$0.07 &  -0.25$\pm$0.03 & 16.75 & \nodata & \nodata & 16.90 \\
J064529.73+412642.0 & 101.373890 &  41.445024 & 55297$\pm$0  & 5.07$\pm$0 &  2.50$\pm$0 & 4.81 & \nodata & \nodata & 16.61 \\
J064558.40+111223.6 & 101.493340 &  11.206556 & 39831$\pm$329 & 5.54$\pm$0.06 &  0.74$\pm$0.08 & 4.66 & 17.44 & 17.52 & 17.68 \\
J070700.65+193526.9 & 106.752730 &  19.590809 & 51530$\pm$469 & 6.50$\pm$0.07 &  1.08$\pm$0.20 & 12.07 & \nodata & \nodata & 17.42 \\
J084223.13+375900.2 & 130.596410 &  37.983394 & 59127$\pm$263  & 6.27$\pm$0.10 &  2.40$\pm$3.53 & 12.98 & 17.04 & 17.51 & 17.86 \\
J084350.85+361419.5 & 130.961886 &  36.238770 & 38293$\pm$166 & 5.62$\pm$0.01 &  2.27$\pm$0.04 & 7.12 & 16.87 & 17.01 & 17.23 \\
J091029.42+090205.1 & 137.622608 &   9.034770 & 39329$\pm$235 & 6.23$\pm$0.02 &  1.50$\pm$0.04 & 5.05 & 17.00 & 17.09 & 17.26 \\
J101457.79+451906.1 & 153.740813 &  45.318381 & 42182$\pm$172 & 5.61$\pm$0.01 &  0.57$\pm$0.05 & 9.99 & 19.64 & 17.69 & 19.01 \\
J124931.17+750727.2 & 192.379883 &  75.124244 & 55650$\pm$2437 & 5.17$\pm$0.15 &  -3.68$\pm$0.73 & 6.04 & \nodata & \nodata & 16.83 \\
J180237.05+032550.1 & 270.654400 &   3.430598 & 44673$\pm$561 & 6.07$\pm$0.13 &  1.05$\pm$0.23 & 6.89 & 17.72 & 25.11 & 17.65 \\
J181325.00+065315.9 & 273.354170 &   6.887751 & 50584$\pm$434 & 5.80$\pm$0.06 &  0.60$\pm$0.07 & 21.92 & \nodata & \nodata & 15.96 \\
J182942.24+110429.1 & 277.426030 &  11.074758 & 52791$\pm$533 & 6.07$\pm$0.07 &  0.60$\pm$0.01 & 16.52 & \nodata & \nodata & 16.70 \\
J205901.96+334933.0 & 314.758180 &  33.825842 & 74995$\pm$7667 & 5.97$\pm$0.27 &  2.14$\pm$11.20 & 4.01 & \nodata & \nodata & 17.49 \\
J210132.70+135622.5 & 315.386270 &  13.939585 & 74981$\pm$251 & 6.50$\pm$0.01 &  -0.33$\pm$0.06 & 29.44 & \nodata & \nodata & 16.49 \\
J230829.86+214333.8 & 347.124442 &  21.726063 & 28470$\pm$536 & 5.07$\pm$0.11 &  -2.86$\pm$0.13 & 3.23 & 17.16 & 15.94 & 15.60 \\

\hline
\end{tabular}
\label{sdOB}
\end{table}

\end{document}